\documentclass[aps,prl,twocolumn,showpacs,amsmath,amssymb,groupedaddress,longbibliography]{revtex4-1}
\usepackage[utf8]{inputenc}
\usepackage{graphicx,graphics}
\usepackage{hyperref}

\hypersetup{
  colorlinks=true,linkcolor=blue,urlcolor=blue,citecolor=blue}
\begin{document}
\title{Tuning Dissipation and Excitations in Superfluid Fermi Gases with a Moving Impurity}
\author{Dong-Chen Zheng}
\author{Yan-Qiang Yu}
\author{Renyuan Liao}\email{ryliao@fjnu.edu.cn}

\affiliation{Fujian Provincial Key Laboratory for Quantum Manipulation and New Energy Materials, College of Physics and Energy, Fujian Normal University, Fuzhou 350117, China}
\affiliation{Fujian Provincial Collaborative Innovation Center for Optoelectronic Semiconductors and Efficient Devices, Xiamen, 361005, China}
\date{\today}
\begin{abstract}
    We develop a method to extract the dissipation for a heavy moving impurity immersed in superfluid Fermi gases. The drag force is derived analytically. As a reward, we are able to extract the dynamical structure factor, from which density excitations of the system is carefully examined. We show that dissipations through drag force is associated with two types of excitations, one being single-particle and the other being collective. We map out the critical velocity for dissipation across the BEC-BCS crossover, consistent with existing experiments. For a magnetic impurity, we show that the dissipation is immune to collective excitations. Our study clearly manifests that dissipation and associated excitations can be controlled by coupling superfluid Fermi gases with a moving impurity, and paves the way for further exploring intriguing realm of nonequilibrium phenomena and dissipation dynamics.
\end{abstract}
\maketitle
Ultracold quantum gases and artificial gauge fields have emerged as an excellent platform to explore many-body systems and to simulate novel phases of matter~\cite{BLO08,GIO08,CHI10,DAL11,RIT13,GOL14,ZHA15,SPI19}. One of the most remarkable achievements is the realization of interacting two-component Fermi gases undergoing the smooth crossover from Bose-Einstein condensates (BECs) to Bardeen-Cooper-Schrieffer (BCS) superfluids~\cite{PIT16}. Universal physical behaviors~\cite{HO04,SAL10,MUK10,ZWI12} emerge at the unitary limit, which represents an strongly interacting system ideal for testing many-body theories~\cite{RAN14,URB18}. Hallmarks of superfluidity such as frictionless flows~\cite{KET07,MOR15,SAL15}, quantized vortices~\cite{KET05} and second sound~\cite{STR13} have been observed in cold atoms.

Collective excitations and dissipation in quantum systems are at the heart of many problems in science and technology~\cite{SAT12,EUG16,CHA18}. There have been intense experimental efforts in probing the collective excitations~\cite{HOI17,BEH18} and dissipations~\cite{KET07,MOR15,SAL15,BUR18} in Fermi superfluids. Theoretical understanding of dissipation of superfluidity dated back to Landau~\cite{LAN41}, who concluded that the critical velocity above which dissipation arises can be related to the elementary excitations of the system through $v_c=min_q[\omega(q)/q]$, with $\omega(q)$ being the spectrum of elementary excitations. At zero temperature, the three-dimensional homogeneous balanced superfluid Fermi gases supports two branches of excitation spectrum~\cite{STR06,KUR16,CAS17,HOI17,ZOU18}: one fermionic branch involving internal degrees of freedom of Cooper pairs and one bosonic branch of excitations of their center of mass motion. Theoretical calculation of critical velocity in Fermi superfluid suggests that it is maximum around unitary~\cite{STR06,HO06,RAN08,STR11}.

Recent advances in cold-atom experiments have enabled a large variety of quantum impurity problem to be explored~\cite{EUG16,JIN16,PAR16,ROA17,KIL18,DEM18}. For a heavy moving impurity in the Bose-Einstein condensates, the drag force experienced by the impurity is calculated by a pioneering work~\cite{PIT04}, and later extended to spin-orbit-coupled Bose-Einstein condensates~\cite{HE14,LIA16}. However, calculating the drag force in the superfluid Fermi gases across the BEC-BCS crossover remains a theoretical challenge, partly due to two reasons: one is the fact that the superfluid order parameter is not directly related to the number density except in the deep BEC limit; the other is that the variation of the order parameter need to be self-consistently taken into account. In this work, we report the first derivation of the drag force from a microscopic theory. This represents an unexpected reward, which enable us to understand the physical mechanism underlying the dissipations. Moreover, it can serve as a theoretical model to understand the recent experimental results where satisfactory theoretical explanation is still lacking~\cite{MOR15}.

We consider a heavy impurity moving with velocity $\mathbf{v}$ in three-dimensional homogeneous two-species atomic Fermi gases interacting via an attractive contact potential, described by the following grand canonical Hamiltonian
\begin{eqnarray}
   H&=&\int d^3\mathbf{r}\sum_{\sigma=\uparrow,\downarrow} \psi_\sigma^\dagger(\mathbf{r}) \left[\frac{\hat{\mathbf{P}}^2}{2m}-\mu+g_I\delta^3(\mathbf{r-v}t)\right]\psi_\sigma(\mathbf{r})\nonumber\\
   &&-g\int d^3\mathbf{r}\psi_\uparrow^\dagger(\mathbf{r})\psi_\downarrow^\dagger(\mathbf{r})\psi_\downarrow(\mathbf{r})\psi_\uparrow(\mathbf{r}).
\end{eqnarray}
Here, $\psi^\dagger_\sigma (\psi_\sigma)$ is the fermionic creation (annhilation) operator for atomic species $\sigma$, $\mu$ is the chemical potential for either species where we have assumed balanced population, and $g_I$ denotes the interaction strength between the impurity and an atom. We consider pairing between different hyperfine species of the same atom, as we restrict ourself to a single mass $m$. For convenience, we shall set $\hbar=2m=1$. The interaction strength $g$ may be expressed in favor of the s-wave scattering length via the prescription: $m/(4\pi a_s)=-1/g+(1/V)\sum_\mathbf{k}m/\mathbf{k}^2$, where $V$ is the volume. We define the Fermi momentum by using $k_F=(3\pi^2n_0)^{1/3}$, with total density $n_0=n_\uparrow+n_\downarrow$, so that the Fermi velocity becomes $v_F=k_F/m$, and the Fermi energy is $E_F=k_F^2/2m$. Throughout this work, we shall work at zero temperature and keep the total density fixed.

 The dynamics of the system can be described by the time-dependent Bogoliubov-deGennes (TDBdG) equations
\begin{eqnarray}
   i\partial_t\begin{bmatrix}u_\nu(\mathbf{r},t) \\ v_\nu(\mathbf{r},t)
   \end{bmatrix}=\begin{bmatrix}\hat{h}(\mathbf{r},t)&\Delta(\mathbf{r},t)\\ \Delta^*(\mathbf{r},t) & -\hat{h}(\mathbf{r},t)
   \end{bmatrix}\begin{bmatrix}u_\nu(\mathbf{r},t) \\ v_\nu(\mathbf{r},t)
   \end{bmatrix},
\end{eqnarray}
where $\hat{h}=\hat{h}_0+\hat{\tilde{h}}$ with $\hat{h}_0=-\nabla^2-\mu$ and $\hat{\tilde{h}}=g_I\delta(\mathbf{r-v}t)$. $u_\nu$ and $v_\nu$ are space- and time- dependent quasiparticle amplitudes satisfying $\int d^3\mathbf{r}u_\nu^*(\mathbf{r},t)u_{\nu^\prime}(\mathbf{r},t)+v_\nu^*(\mathbf{r},t)v_{\nu^\prime}(\mathbf{r},t)=\delta_{\nu\nu^\prime}$. In general, the above equations must be solved together with the equation for the order parameter $\Delta(\mathbf{r},t)=-g\sum_{\nu}u_\nu(\mathbf{r},t) v_{\nu}^*(\mathbf{r},t)$ and the equation for the number density $n(\mathbf{r},t)=2\sum_\nu|v_\nu(\mathbf{r},t)|^2$. However such calculation is numerically very demanding. In this work, we shall adopt another approach which is reasonable and theoretically transparent when the interaction strength $g_I$ between the impurity and an atom  is assumed to be weak.

To proceed, we write $u_\nu(\mathbf{r},t)=u_\nu(\xi)e^{-i\epsilon_\nu t }$ and $v_\nu(\mathbf{r},t)=v_\nu(\xi)e^{-i\epsilon_\nu t}$, where $\xi=\mathbf{r-v}t$. After the substitution into the TDBdG equations, we end up with an eigenvalue problem
\begin{eqnarray}
   \begin{bmatrix}
      \hat{h}+i\mathbf{v}\cdot\nabla &  \Delta(\xi)\\
      \Delta^*(\xi) &  -\hat{h}+i\mathbf{v}\cdot\nabla
   \end{bmatrix}\begin{bmatrix}u_\nu \\ v_\nu
   \end{bmatrix}=\epsilon_\nu\begin{bmatrix}u_\nu\\ v_\nu \end{bmatrix}.
\end{eqnarray}
To linearize the equations, we make the following decompositions: $\Delta=\Delta_0+\tilde{\Delta}$, $u_\nu(\xi)=u_\mathbf{p}e^{i\mathbf{p\cdot\xi}}+\tilde{u}_\mathbf{p}(\xi)$ and $v_\nu(\xi)=v_\mathbf{p}e^{i\mathbf{p\cdot\xi}}+\tilde{v}_\mathbf{p}(\xi)$, where $u_\mathbf{p}$ and $v_\mathbf{p}$ are the solutions when $g_I$ vanishes. We make Fourier transformations $\tilde{u}_\mathbf{p}(\xi)=\sum_\mathbf{k}\tilde{u}_{\mathbf{pk}}e^{i\mathbf{k\cdot\xi}}$, $\tilde{v}_\mathbf{p}(\xi)=\sum_\mathbf{k}\tilde{v}_\mathbf{pk}e^{i\mathbf{k\cdot\xi}}$, and $\tilde{\Delta}=\sum_\mathbf{k}\tilde{\Delta}_\mathbf{k}e^{i\mathbf{k\cdot\xi}}$. The linearized equations for the fluctuating parts $\tilde{u}_{\mathbf{pk}}$ and $\tilde{v}_{\mathbf{pk}}$ are as follows
\begin{eqnarray}
    \mathcal{G}^{-1}(\mathbf{p},\mathbf{k})\begin{bmatrix}\tilde{u}_{\mathbf{pk}}\\ \tilde{v}_{\mathbf{pk}}\end{bmatrix}=\begin{bmatrix} g_I &\tilde{\Delta}_{\mathbf{k-p}} \\ \tilde{\Delta}_{\mathbf{p-k}}^*&-g_I\end{bmatrix}\begin{bmatrix}u_\mathbf{p}\\ v_\mathbf{p}\end{bmatrix},
    \label{eq:eigfun}
   \end{eqnarray}
where we have defined the matrix $\mathcal{G}^{-1}(\mathbf{p},\mathbf{k}) $ as
\begin{eqnarray}
   \mathcal{G}^{-1}(\mathbf{p},\mathbf{k})=\begin{pmatrix}-\xi_{\mathbf{k}}+\mathbf{k}\cdot\mathbf{v}+\epsilon_\mathbf{p} & -\Delta_0\\
   -\Delta_0 & \xi_\mathbf{k}+\mathbf{k\cdot v}+\epsilon_\mathbf{p}\end{pmatrix}.
\end{eqnarray}\label{eq:five}
 In the above, $\xi_\mathbf{k}=\mathbf{k}^2-\mu$, and $\Delta_0$ (we choose to be real due to gauge degree of freedom) is the order parameter in the absence of the impurity. It should be noted that the pole of $\mathcal{G}^{-1}$ determines the eigenenergy of the unperturbed system in the moving frame  $\epsilon_\mathbf{p}$=$E_\mathbf{p}-\mathbf{p\cdot v}$ with $E_\mathbf{p}=\sqrt{\xi_\mathbf{p}^2+\Delta_0^2}$. The corresponding eigenfunction is given by $[u_\mathbf{p},v_\mathbf{p}]^T=[\sqrt{(1+\xi_\mathbf{p}/E_\mathbf{p})/2},\sqrt{(1-\xi_\mathbf{p}/E_\mathbf{p})/2}]^T$.

 The number density  and the order parameter can be evaluated to linear order in fluctuations of eigenfunctions:
 \begin{subequations}
 \begin{eqnarray}
      n(\xi)&=&2\sum_{\mathbf{p}}|v_\mathbf{p}e^{i\mathbf{p\cdot\xi}}+\tilde{v}_\mathbf{p}|^2\nonumber\\
      &\approx&2\sum_\mathbf{p}v_\mathbf{p}^2+2\sum_\mathbf{p}(v_\mathbf{p}e^{i\mathbf{p\cdot\xi}}\tilde{v}_\mathbf{p}^*+c.c)\nonumber\\
      &=&n_0+\delta{n}(\xi),\\
      \Delta(\xi)&=&-g\sum_\mathbf{p}(u_\mathbf{p}e^{i\mathbf{p\cdot \xi}}+\tilde{u}_\mathbf{p})(v_\mathbf{p}e^{i\mathbf{p\cdot\xi}}+\tilde{v}_\mathbf{p})^*\nonumber\\
      &\approx&-g\sum_\mathbf{p}u_\mathbf{p}v_\mathbf{p}-g\sum_\mathbf{p}(u_\mathbf{p}e^{i\mathbf{p\cdot\xi}}\tilde{v}_\mathbf{p}^*+c.c)\nonumber\\
      &=&\Delta_0+\tilde{\Delta}(\xi).
 \end{eqnarray}
 \label{eq:number}
 \end{subequations}
It should be pointed out that variations of the density is correlated with the variation of the order parameter through the fluctuations of the eigenfunctions, as could be seen from equations ($\ref{eq:eigfun}$) and ($\ref{eq:number}$).
After lengthy and sophisticated manipulations (for details, please see~\cite{SM19}), we finally obtain the drag force experienced by the impurity
\begin{eqnarray}
   \mathbf{F}&=&-\int d^3\mathbf{r}\vec{\nabla} [g_I\delta^3(\mathbf{r-v}t)] n(\mathbf{r-v}t)\nonumber\\
   &=&g_I\vec{\nabla}\delta n(\mathbf{r-v}t)|_{\mathbf{r}=\mathbf{v}t}\nonumber\\
   &=&2g_I^2\sum_\mathbf{q}i\mathbf{q}\mathcal{D}(\mathbf{q},iw_m\rightarrow \mathbf{q\cdot v}+i0^+).
   \label{eq:force}
\end{eqnarray}
Several remarks are in order. The infinitesimal imaginary part was added following the usual causality rule~\cite{PIT04,CAR06,LIA16}. The connection of the drag force with the dynamical structure factor is a manifestation of fluctuation-dissipation theorem~\cite{SIM10}. The dynamical structure factor $\mathcal{D}(\mathbf{q},z)$ reads
\begin{eqnarray}
  \mathcal{D}(\mathbf{q},z)&=&\mathcal{D}_{pb}-\Delta_0^2\frac{I_{11}A^2+z^2I_{22}B^2-2z^2I_{12}AB}{I_{11}I_{22}-z^2I_{12}^2}\nonumber\\
  &\equiv&\mathcal{D}_{pb}(\mathbf{q},z)+\mathcal{D}_{cl}(\mathbf{q},z),
\end{eqnarray}
where we have defined
\begin{subequations}
\begin{eqnarray}
    A(\mathbf{q},z)&=&\sum_\mathbf{p}\frac{E_++E_-}{2E_+E_-}\frac{\xi_++\xi_-}{z^2-(E_++E_-)^2},\\
    B(\mathbf{q},z)&=&\sum_\mathbf{p}\frac{E_++E_-}{2E_+E_-}\frac{1}{z^2-(E_++E_-)^2},\\
    \mathcal{D}_{pb}(\mathbf{q},z)&=&\sum_\mathbf{p}\frac{E_++E_-}{2E_+E_-}\frac{E_+E_--\xi_+\xi_-+\Delta_0^2}{z^2-(E_++E_-)^2}.
\end{eqnarray}
\end{subequations}
and
 \begin{eqnarray}
    I_{11}&=&\sum_\mathbf{p}\frac{E_++E_-}{2E_+E_-}\frac{E_+E_-+\xi_+\xi_-+\Delta_0^2}{z^2-(E_++E_-)^2}+\frac{1}{2E_\mathbf{p}},\nonumber\\
    I_{22}&=&\sum_\mathbf{p}\frac{E_++E_-}{2E_+E_-}\frac{E_+E_-+\xi_+\xi_--\Delta_0^2}{z^2-(E_++E_-)^2}+\frac{1}{2E_\mathbf{p}},\nonumber\\
    I_{12}&=&\sum_\mathbf{p}\frac{1}{2E_+E_-}\frac{E_+\xi_-+E_-\xi_+}{z^2-(E_++E_-)^2}.
 \end{eqnarray}
 with $\pm$ being a shorthand notation for momentum $\mathbf{p\pm q/2}$. Identical expressions for the dynamical structure factor within the BCS mean-field theory have been obtained by  various approaches, including kinetic equations~\cite{STR06,GUO13}, the random-phase approximation~\cite{GRI93,ZOU10} and functional integrals~\cite{HE16}. Measurements of the dynamical structure factor~\cite{HOI17,HUL18} via two-photon Bragg spectroscopy~\cite{KET99,ZAM01} performed recently reveal salient physics.

 The drag force can be decomposed into two parts: $\mathbf{F}=\mathbf{F}_{pb}+\mathbf{F}_{cl}$, with one part $\mathbf{F}_{pb}=-g_I^2\sum_\mathbf{q}i\mathbf{q}\mathcal{D}_{pb}(\mathbf{q},iw_m\rightarrow \mathbf{q\cdot v}+i0^\dagger)$ being the contribution from pair-breaking excitations and the other one $\mathbf{F}_{cl}=-g_I^2\sum_\mathbf{q}i\mathbf{q}\mathcal{D}_{cl}(\mathbf{q},iw_m\rightarrow\mathbf{q\cdot v}+i0^\dagger)$ from collective mode excitations. The pair-breaking excitation spectrum $\omega_{pb}$  corresponds to the pole of $\mathcal{D}_{pb}(\mathbf{q},z)$, namely $\omega_{pb}=E_++E_-$. It is a single-particle continuum, and its minimum $\omega_{th}(q)$ denotes the threshold energy to breaking a Cooper pair with center of mass momentum $\mathbf{q}$. The collective spectrum $\omega(\mathbf{q})$ can be found by seeking the poles of $\mathcal{D}_{cl}(\mathbf{q},z)$, yielding
 \begin{eqnarray}
    I_{11}(\mathbf{q},\omega)I_{22}(\mathbf{q},\omega)-\omega^2I_{12}^2(\mathbf{q},\omega)=0.
 \end{eqnarray}
\begin{figure}[t]
\includegraphics[width=1\columnwidth,height=0.60\columnwidth]{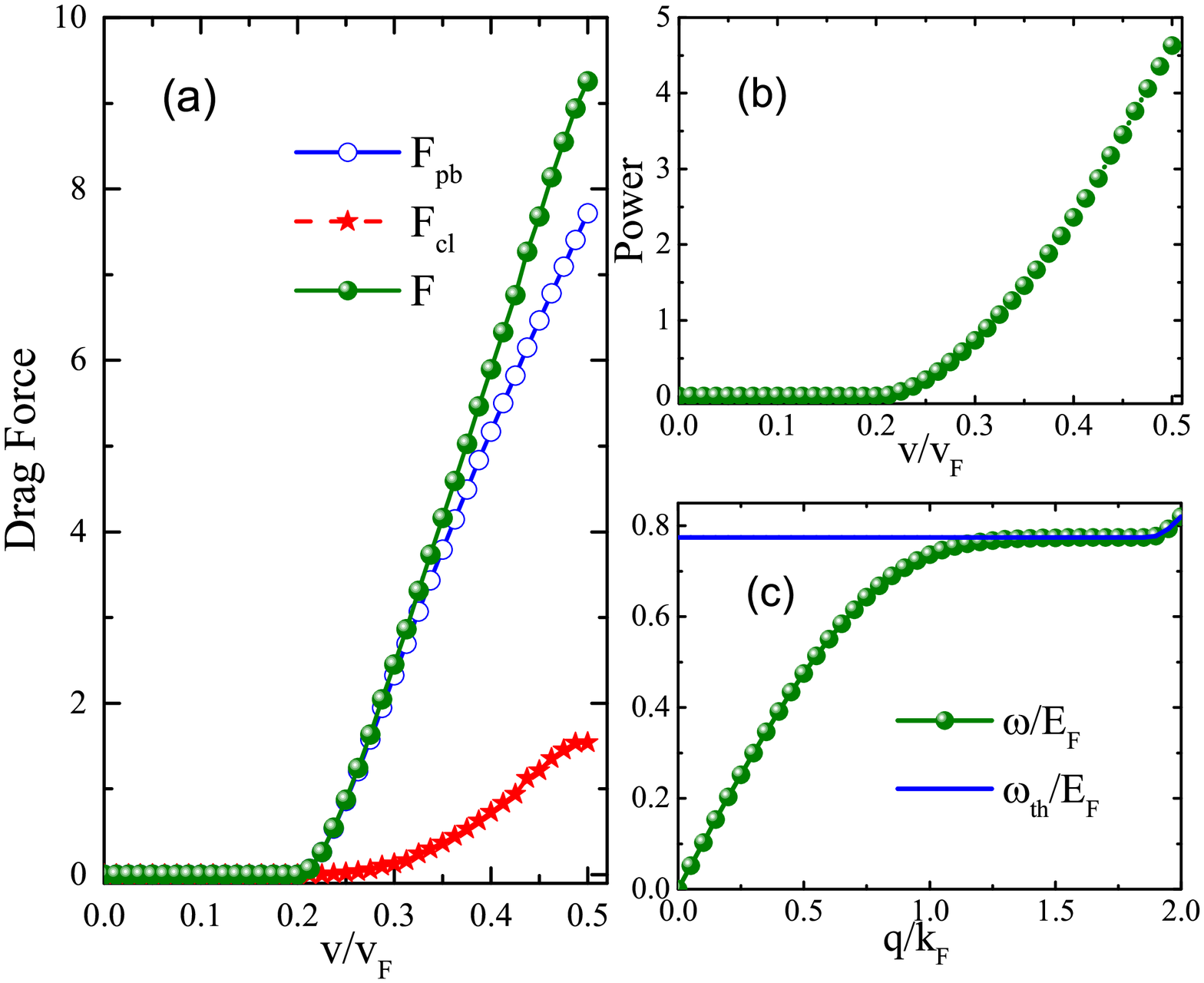}
\caption{(color online) Dissipation and excitations in the BCS regime where $1/k_Fa_s=-0.5$: (a) the total drag force F and its two contributions $F_{pb}$ and $F_{cl}$, in units of $g_I^2n_0^2/k_F$; (b) the power absorbed by the system, in units of $g_I^2n_0^2$; (c) the collective excitation $\omega$ and the threshold energy  $\omega_{th}$ for pair-breaking excitation.  Here the total drag force $F=F_{pb}+F_{cl}$ has two contributions. $F_{pb}$ has origin from pair-breaking excitations and $F_{cl}$ has origin from collective excitations. The concavity of the collective spectrum renders that the sound velocity sets an upper bound for collective excitations. }
\label{fig1}
\end{figure}
 On the BCS side, as shown in Fig.$\ref{fig1}$, the drag force $F$ develops a nonzero value when the velocity $v$ exceeds the pair-breaking velocity $v_{pb}=min_\mathbf{q}(E_++E_-)/q=0.203V_F$. At this stage, the drag force has the contribution solely from the pair-breaking excitations with $F=F_{pb}$ and $F_{cl}=0$. When the velocity is further increased to $v=0.247v_F$, $F_{cl}$ starts to increase from zero. As shown in panel (b), the power ($P=Fv$) absorbed by the system increases monotonically with the velocity, following the same trend of the total drag force. The sound velocity can be determined via $v_s=lim_{q\rightarrow 0}\omega/q=0.501v_F$. Remarkably, dissipation due to collective excitations emerges when  the velocity reaches only half of the sound velocity. In other words, the sound velocity sets an upper bound for dissipation from collective excitations. This is attributed to the concavity of the collective excitation branch on the BCS side~\cite{CAS15,TOI15,KUR16}, shown in panel (c), where the slope of the collective spectrum decreases gradually.

\begin{figure}
\includegraphics[width=1\columnwidth,height=0.60\columnwidth]{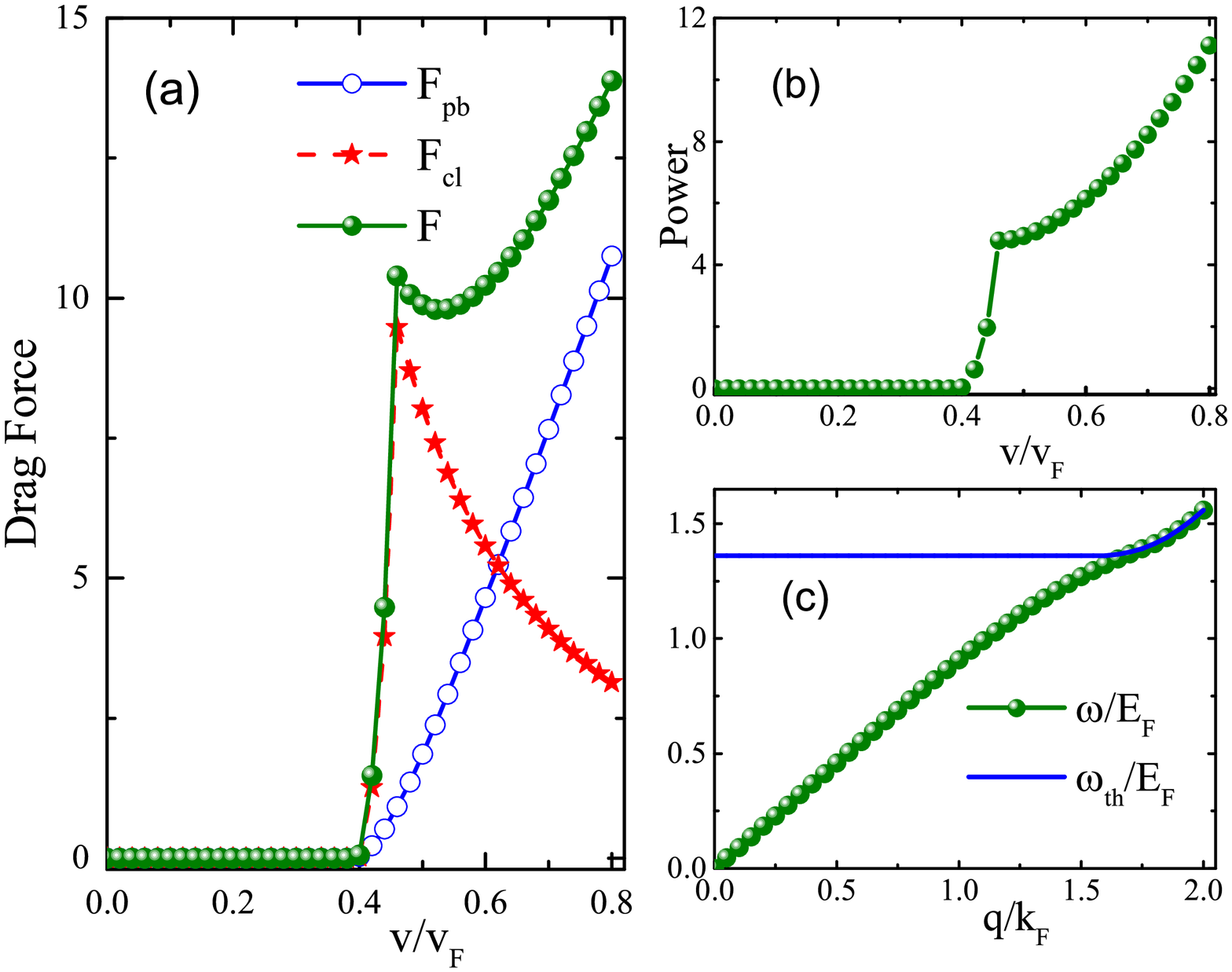}
\caption{(color online) Dissipation and excitations in the unitary limit where $1/k_Fa_s=0$: (a) the drag force $F$, $F_{pb}$ and $F_{cl}$, in units of $g_I^2n_0^2/k_F$; (b) the power absorbed by the system, in units of $g_I^2n_0^2$ ; (c)the spectrum of collective excitations $\omega$ and the threshold energy $\omega_{th}$ for pair-breaking excitations. There appears a local maximum of the total drag force, resulting from the cooperative effects of single-particle and collective excitations. The almost linear behavior of the collective spectrum indicates that sound velocity faithfully reflects the threshold for the collective excitations.}
\label{fig2}
\end{figure}

 Let's turn to unitary limit, which is both theoretically intriguing and experimentally interesting. As shown in Fig.~$\ref{fig2}a$, it seems that $F_{pb}$ and $F_{cl}$ appears nonzero almost at the same threshold velocity. However, a close inspection indicates that the onset of $F_{pb}$ starts as the velocity reaches pair-breaking velocity $v_{pb}=0.390v_F$ while the onset of $F_{cl}$ occurs as long as the velocity approaches the sound velocity $v_s=0.408v_F$. This may be explained by the noticeable feature manifested by the dispersion shown in panel (c): it is surprisingly almost linear up to the merging with the continuum. Interestingly, while $F_{pb}$ increases monotonically with the velocity, $F_{cl}$ increases sharply when the velocity exceeds the sound velocity before it reaches a local maximum around $v=0.465v_F$ and decreases as the velocity increases further. This results in a non-monotonic behavior for the total force $F$. Despite the peculiar behavior of $F$, the power absorbed by the system still increases with the velocity, manifesting a good indicator for dissipation.

\begin{figure}
\includegraphics[width=1\columnwidth,height=0.60\columnwidth]{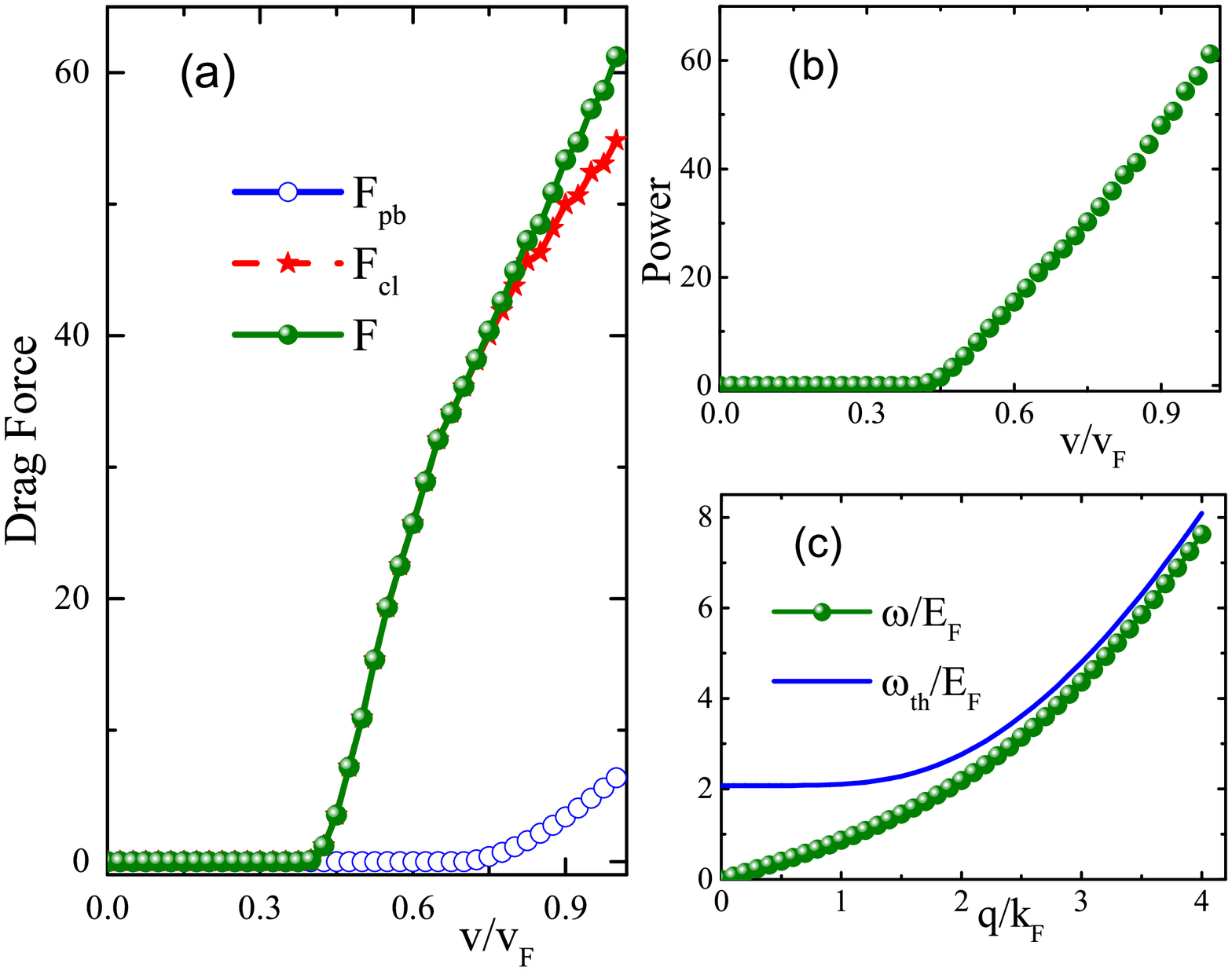}
\caption{(color online) Dissipation and excitations in the BEC regime where $1/k_Fa_s=0.5$: (a) the drag force $F$, $F_{pb}$ and $F_{cl}$, in units of $g_I^2n_0^2/k_F$; (b) the power absorbed by the system, in units of $g_I^2n_0^2$ ; (c)the collective excitation $\omega$ and the threshold energy $\omega_{th}$ for pair-breaking excitations. The low-lying excitations is dominated by collective excitations. The convexity of the collective spectrum suggests that the sound velocity sets a lower bound for collective excitation.}
\label{fig3}
\end{figure}

 Now come to the physics on the BEC side where $1/k_Fa_s=0.5$. At low energy, the internal degrees of freedom for the Cooper pairs get frozen out, and the physics involving collective excitations becomes dominant. The pair-breaking threshold always lies higher than the collective spectrum, as shown on panel (c) in Fig.~\ref{fig3}. The convex nature of the collective spectrum suggests that sound velocity sets a lower bound for dissipation. The drag force appears when the velocity exceeds the sound velocity $v_s=0.385v_F$. When the velocity exceeds the pair-breaking velocity $v_{pb}=0.690v_F$, $F_{cl}$ starts to appear and contributes to a larger dissipation, as clearly shown in panel (b), where the power absorbed by the system increases somehow linearly with the velocity.

\begin{figure}
\includegraphics[width=1\columnwidth,height=0.60\columnwidth]{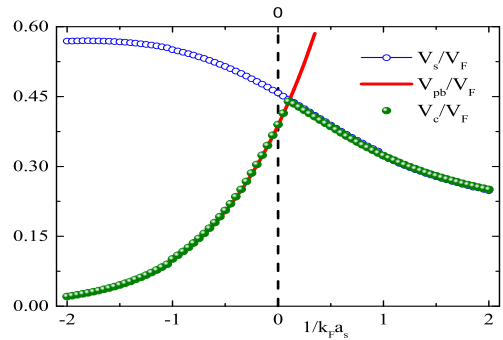}
\caption{(color online) The sound velocity $v_s$, the pair-breaking velocity $v_{pb}$, and the critical velocity $v_c$ across BEC-BCS crossover. Here $v_c$ is determined by the threshold velocity at which the drag force starts to emerge. The pair-breaking velocity $v_{pb}$ increases monotonically along the crossover into the BEC regime.  The sound velocity is determined via $v_s=lim_{q\rightarrow 0}\omega(q)/q$, corresponding to the Anderson-Bogoliubov mode in the BCS limit with $v_s=v_F/\sqrt{3}$, expected from broken symmetry of superfluid phase. The vertical dash line is plotted for better vision.}
\label{fig4}
\end{figure}

Critical velocity $v_c$ is an important quantity for superfluids characterizing the threshold velocity above which dissipation arises. In our situation, this corresponds to the velocity driving the onset of the total drag force $F$. We show the critical velocity $v_c$, the pair-breaking velocity $v_{pb}$, and the sound velocity $v_s$ in Fig.~$\ref{fig4}$. For weakly bound fermions, the critical velocity is proportional to the binding energy of the pairs, which increases monotonically along the crossover into the BEC regime. The sound velocity $v_s$ which sets the critical velocity for phonon excitation, decreases monotonically from BCS to BEC side, where in the BCS limit it approaches the Anderson-Bogoliubov mode with $v_s=v_F/\sqrt{3}$ and in the BEC limit it becomes $\sqrt{k_Fa_s/3\pi}v_F$, as the system can be regarded as weakly-interacting Bose-Einstein condensates of diatomic molecules with mass $2m$, density $n_0/2$ and inter-molecular scattering length $a_M=2a_s$~\cite{PIE03}. The critical velocity we determined nicely follows the minimum of $v_{pb}$ and $v_s$, showing a pronounced peak around the unitary, consistent with the experimental results~\cite{KET07,MOR15}.

So far we have focused on a non-magnetic impurity. A magnetic impurity provides interesting probing to a cold atomic system~\cite{PU11,JAK16}. The theoretical formulation we developed above can be conveniently generalized for a magnetic impurity with spin-dependent impurity-atom coupling $g_I^\sigma=\sigma _zg_I$. It turns out that the drag force is found to be (for details, please see~\cite{SM19})
\begin{eqnarray}
    \mathbf{F}_s=2g_I^2\sum_\mathbf{q}i\mathbf{q} \mathcal{D}_s(\mathbf{q},iw_m\rightarrow \mathbf{q\cdot v}+i0^\dagger),
\end{eqnarray}
where the dynamical spin structure factor is given by
\begin{eqnarray}
\mathcal{D}_s(\mathbf{q},z)=\sum_\mathbf{p}\frac{E_++E_-}{2E_+E_-}\frac{\xi_+\xi_-+\Delta_0^2-E_+E_-}{z^2-(E_++E_-)^2}.
\end{eqnarray}
\begin{figure}
\includegraphics[width=1\columnwidth,height=0.60\columnwidth]{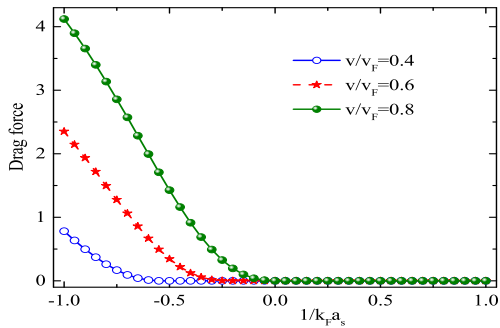}
\caption{(color online) The drag force for a magnetic impurity at different velocities across the BEC-BCS crossover. Only single-particle excitation contributes to the drag force since the magnetic impurity probes the internal degrees of freedom of the pairs, leaving the collective excitations intact.}
\label{fig5}
\end{figure}
Examining the pole structure of $D_s$ indicates that only single-particle excitation is involved in the drag force. This is reasonable because a magnetic impurity probes the internal degrees of freedom of Cooper pairs instead of collective excitations associated with the center of mass motion of the pairs. The drag force for a magnetic impurity across the BEC-BCS crossover for some fixed velocities is shown in Fig.~$\ref{fig5}$. For a given velocity, the drag force becomes more prominent as the system is tuned toward the BCS side, as the magnetic impurity only probes pair-breaking excitations. Increasing the velocity leads to the rising of the drag force and enhanced dissipation.

In summary, We derived an analytical expression for the drag force experienced by a moving impurity. This enables us to understand the underlying mechanism responsible for dissipation. Our work is expected to pave the way for a better understanding of superfluidity and dissipation dynamics in the intriguing regime of nonequilibrium physics.

%

 We acknowledge stimulating discussions with Lin Wen. This work is supported by NSFC under Grant No. $11674058$ and $11891240378$.

\end{document}